

\magnification=\magstep1
\hsize=13cm
\vsize=20cm
\overfullrule 0pt
\baselineskip=13pt plus1pt minus1pt
\lineskip=3.5pt plus1pt minus1pt
\lineskiplimit=3.5pt
\parskip=4pt plus1pt minus4pt

\def\negenspace{\kern-1.1em}



\newcount\secno
\secno=0
\newcount\susecno
\newcount\fmno\def\z{\global\advance\fmno by 1 \the\secno.
                       \the\susecno.\the\fmno}
\def\section#1{\global\advance\secno by 1
                \susecno=0 \fmno=0
                \centerline{\bf \the\secno. #1}\par}
\def\subsection#1{\medbreak\global\advance\susecno by 1
                  \fmno=0
       \noindent{\the\secno.\the\susecno. {\it #1}}\noindent}


\def\sqr#1#2{{\vcenter{\hrule height.#2pt\hbox{\vrule width.#2pt
height#1pt \kern#1pt \vrule width.#2pt}\hrule height.#2pt}}}


\newcount\refno
\refno=1
\def\y{\the\refno}
\def\myfoot#1{\footnote{$^{(\y)}$}{#1}
                 \advance\refno by 1}


\def\neq{\hbox{$\,$=\kern-6.5pt /$\,$}}





\newcount\secno
\secno=0
\newcount\fmno\def\z{\global\advance\fmno by 1 \the\secno.
                       \the\fmno}
\def\sectio#1{\medbreak\global\advance\secno by 1
                  \fmno=0
       \noindent{\the\secno. {\it #1}}\noindent}





\magnification=\magstep1
\hsize 13cm
\vsize 20cm
\hfill{Preprint IMAFF 94/7}
\bigskip
\centerline{\bf{NON-LINEAR AFFINE EMBEDDING OF THE DIRAC FIELD}}
\centerline{\bf{ FROM THE
MULTIPLICITY-FREE SL(4,R) UNIRREPS}}

\vskip 1.0cm
\centerline{by}
\vskip 1.0cm
\centerline{A. L\'opez--Pinto, A. Tiemblo and R. Tresguerres}
\vskip 1.0cm
\centerline{\it {IMAFF, Consejo Superior de Investigaciones Cient¡ficas,}}
\centerline{\it {Serrano 123, Madrid 28006, Spain}}
\vskip 1.5cm
\centerline{ABSTRACT}\bigskip

The correspondence between the linear multipilcity-free unirreps
of SL(4, R) studied by Ne'eman and {\~{S}}ija{\~{c}}ki and the
non-linear realizations of the affine group is worked out. The
results obtained clarify the inclusion of
spinorial fields in a non-linear affine gauge theory of gravitation.\bigskip
 \bigskip\bigskip

\sectio{\bf{Introduction}}\bigskip
The metric affine gauge theories (MAGTs) started by Hehl {\it
et al} (for a valuable review see [1]) provide a powerful
mathematical framework for the study of gravitation. They
enable to deal with a geometrical theory endowed with the
most general connection and based on the most general
transformation group
 in the tangent space, including all the possible gauge degrees
of freedom. Furthermore, it has been
suggested that the adoption of the affine group as the dynamical
group in a gauge theory of gravitation might solve certain
renormalizability and unitarity problems in quantum
gravity$^{[1]}$.\bigskip
Nevertheless, in [2] and [3] we showed the close relationship between
the standard MAGTs and the non-linear affine gauge theories of
gravitation (NLAGTs). In the context of a NLAGT, the matter fields
appear as a sector of the representation space  of the
non-linear realization. They vary as a linear representation
of the Lorentz group with coefficients depending on certain fields
(coset fields) and on the element of the affine group involved
(non-linearity, see below).
\bigskip
There is a strong parallelism between the objects in a linear
gauge theory and those in the non-linear version. This link
was studied by us for the case of the affine group pointing out the
relationship between most of the objects, as connections and
curvatures is easily established. However, when we deal with the
matter fields, the relationship of the linear and non-linear
objects presents serious difficulties. We can always start with
a given linear field and study the non-linear one that it
induces. The reverse problem is more complicated. It might be expected
 that the reducible non-linear field is obtained from a linear
representation of the gauge group. We then need a linear
field which, when restricted to the action of the
subgroup, breaks into the desired representation of the subgroup
plus other pieces. Nevertheless, this is a hard problem of group
representation theory. This difficulties
can be overpassed, as we have done in our forementioned papers, taking into
account the fact that the non-linear realization is
well defined by itself, and we can work with it ignoring any
correspondence with
a linear representation. However, one can feel tempted to tighten
the relation between the linear and non-linear approaches and
try to obtain the linear representation which induces a given
non-linear one. In this context, if we want to
consider an electron in a gauge version of gravity like the
NLAGT, we should first know a linear representation of the affine
group which contains the Dirac field in its "Lorentz
breaking".\bigskip
In this work, the explicit inclusion of fermionic matter in the
NLAGT, starting from a linear infinite-dimensional
representation of the affine group, is worked out. In section 2
the machinery of the
non-linear$^{[5],[6],[7]}$ realizations is revised with special care to its
application to the NLAGT. The multipilcity-free
unirreps of SL(4,R)$^{[11],[12],[13],[14]}$ and their physical constraints are
presented in section 3. In section 4 use is made of the results of sections 2
and 3 to study the role of the Dirac field in the NLAGT. Finally
the most outstanding conclusions are drawn.\bigskip
\bigskip

\sectio{\bf{Nonlinear realization of the affine group }}\bigskip

The relevance of the nonlinear realization
techniques in different branches of theoretical physics  is
widely known, ranging from
spontaneous symmetry breaking and nonlinear Lagrangians to gauge
versions of space-time
groups (see, for
instance, [4] and [5] and references therein). The
classification of all the nonlinear
realizations of a group which become linear under the action of
a given subgroup H of G was first established by Coleman, Wess
and Zumino$^{[6]}$. They essentially showed that, under several
natural hypothesis, it is always possible to express any
nonlinear realization in what they called the "standard form",
which is nothing but a particular parametrization of the coset
representation. However, we shall follow an approach due to Salam and
Strathdee$^{[7]}$ which becomes specially transparent when dealing with
definite physical problems. Here we give a short review of the scheme.

\bigskip
Let us suppose we know the action of certain irreducible linear
representations of a group G

$$\varphi\longrightarrow
\varphi^{\prime}=\;D\,(g)\;\varphi\quad\,,\qquad g\in
G \eqno(\z)$$

\noindent as well as their breaking into linear irreducible
representations of a subgroup H. We shall make the assumption
that the space basis is chosen in such a way that the matrices $D\,(h)$,
where  $h\epsilon\,H$, break into diagonal boxes.

We define the reduction matrix $(\;L_\sigma)_{\alpha\beta}$
to be the one with elements written in terms of several field
variables with the following restrictions:

 a)$\,\,$ $L_\sigma$ belongs to the group G,
so that for every representation
$g \rightarrow D\,(g)$, the functional $D(L_\sigma)$ is well
defined. Thus, the number of fields $\sigma_a$ needed to
 parametrize $L_\sigma$ is smaller or equals the dimension of G.

 b)$\,\,$Under the group operations, the fields which
parametrize the reducing matrix transform in such a way that

$$L_\sigma\longrightarrow L_{\sigma ^\prime}=\,g\;L_\sigma
\;h^{-1}\;(\sigma ,g)\quad\,,g \in G,\quad\,\, h(\sigma,g)\in H,\eqno(\z)$$

\noindent holds.

\noindent c)$\,\,$When we restrict the operations to the subgroup H, the
reduction matrix varies as usual, namely
$$ L_{\sigma ^\prime}=\,h\;L_\sigma\;h^{-1}.\eqno(\z)$$

\noindent From (2.2) we obtain that  the
functional $D(L_\sigma)$ transforms as

$$D(L_\sigma)\longrightarrow D(L_{\sigma ^\prime})=\,D(g\;L_\sigma
\;h^{-1}\;(\sigma ,g))=\;D(g)\;D(\;L_\sigma)\;D(h^{-1}\;(\sigma
,g)).\eqno(\z)$$

\noindent This property enables us to project the linear
realizations (2.1) into nonlinear ones. If we define

$$ \psi :=\;D({L_\sigma}^{-1})\;\varphi,\eqno(\z)$$

\noindent it can be easily seen, using the expressions (2.1) and
(2.4), that the transformation law for the new fields  $\psi$
is given by
$$ \psi^{\prime}=\;D(h)\; \psi.\eqno(\z)$$

\noindent Hence, $\psi$ transforms under G as a reducible
representation of H with nonlinear parameters $h(\sigma,g)$.
In fact, this formula plays a fundamental role in the subsequent
developments. It makes possible to establish a one to one
relationship between the linear and the non-linear objects. As
it was shown in [2] and [3], (2.5) is not essential for
introducing ordinary matter in a non-linear gauge theory of the
affine group. Nevertheless, the deep correspondence between the
linear and the non-linear theories is better understood if we
find the linear partner for each non-linear object . The main
purpose of this paper will be to state the link between the non-linear
Dirac field and its linear partner.\bigskip
It is important to realize that the space where the
nonlinear realization takes place is spanned not only by the
fields $\psi$ but by the $\sigma^{\prime }$s as well, so that
the group action is given by both (2.2) and (2.6). The "standard
form" of [6] corresponds to

$$L_\sigma:=\;e^{\sigma A},\eqno(\z)$$

\noindent where A are those generators of G which are not
contained in the algebra of H. In this scheme the fields
$\sigma$ are usually called coset parameters.\bigskip

Our purpose is to use the nonlinear realizations as the
mathematical machinery for a physical gauge theory. So, we have
to consider the local realization of the group G, i.e. the group
action when  $g=\,g(x)\in$G where x are the space-time
labels.

 It is easily seen from (2.2) and (2.6) that the
derivative of $\psi$ does not transform covariantly. To this
end, we need to define a covariant derivative with the
desired transformation properties, introducing the algebra-valued one-form

$$\Omega=\,\Omega_{i}\,s_i, \eqno(\z)$$

\noindent where $\{s_i\}_{i=1}^{dim G}$ is a basis of the algebra
of G. The transformation of $\Omega$ under G is
given by

$$\Omega\longrightarrow g\,\Omega\,g^{-1}+\,\,g\,d\,
g^{-1}.  \eqno(\z)$$

\noindent In the gauge version of a linearly realized group G,
$\Omega$ is taken as the physical gauge connection, but this is not the case
now. We define a modified gauge field

$$\Gamma =\,{L_\sigma}^{-1}\,(\Omega +\,\,d)
\,{L_\sigma}.\eqno(\z)$$

\noindent From (2.2) and (2.9) it is easy to obtain the
transformation of $\Gamma$

$$\Gamma\longrightarrow h\,\Gamma\,h^{-1}+\,\,
h\,d\,h^{-1},\eqno(\z)$$

\noindent with h given by (2.2). A very important feature of
 (2.11) is that the homogeneous piece of the transformation
projects only over the algebra of H, so that the components of
$\Gamma$ associated to the generators of G laying outside the
algebra of H, transform "tensorially", i.e. their variation is
homogeneous, the second term in the r.h.s. of (2.11) being absent.

\noindent We now define a covariant differential ${\bf D}$ of the
nonlinear field $\psi$ with the desired transformation properties
as
$${\bf D}\, \psi:=\,(\,d+
i\,\Gamma_ { i}\,S_i)\,\psi,\eqno(\z)$$
\noindent where $S_i$ are the representatives of the $s_i$
generators in the D representation.
\bigskip
Now, we will apply the above given non-linear tools to the
affine group non-linearly realized, with the Lorentz one playing
the role of the classification subgroup H. The results given below
are a short summary of yhe NLGAT approach. \bigskip
 The affine group is the semidirect product of the translations
and the general linear transformations in four dimensions. We
can choose as a basis of the affine algebra the six Lorentz generators
 $L ^\alpha
{}_\beta $, ten pseudo-symmetric linear generators $S ^\alpha
{}_\beta $ and the four translational generators $P_\alpha $.
They fulfill the following conmutation relations:
$$\eqalign{\left[L_{\alpha\beta }\,,L_{\mu\nu }\right]=
-i\,&\left( o_{\alpha [\mu } L_{\nu ]\beta}
         - o_{\beta [\mu }L_{\nu ]\alpha }\right)\,,\cr
\left[L_{\alpha\beta }\,,S_{\mu\nu }\right]=
\,\,\,\,i\,&\left( o_{\alpha (\mu } S_{\nu )\beta}
         - o_{\beta (\mu }S_{\nu )\alpha }\right)\,,\cr
\left[S_{\alpha\beta }\,,S_{\mu\nu }\right]=
\,\,\,\,i\,&\left( o_{\alpha (\mu } L_{\nu )\beta}
         + o_{\beta (\mu }L_{\nu )\alpha }\right)\,,\cr
\left[L_{\alpha\beta }\,, P_\mu\right]\hskip0.10cm=
\,\,\,\,i\,&o_{\mu [\alpha }P_{\beta ]}\,,\cr
\left[S_{\alpha\beta }\,, P_\mu\right]\hskip0.10cm=
\,\,\,\,i\,&o_{\mu (\alpha }P_{\beta )}\,,\cr
\left[P_\alpha\,, P_\beta\right]\hskip0.30cm
=\,\,\,\,\,\,&\,0\,,\cr}\eqno(\z)$$
\noindent where $o_{\alpha\mu}$ is the flat Minkowski metric.

\noindent In the
scheme developed in [2] and [3], the reduction matrix is
$$L_{(\xi,\delta)}:=e^{-i\,\xi ^\alpha
P_\alpha}e^{i\,\delta^{\mu\nu } S_{\mu\nu}}\,,\eqno(\z)$$
\noindent and the analogous  of transformation (2.2) is now

$$g\,L_{(\xi,\delta)}=\,L_{({\xi}^\prime,{\delta}^\prime)}
\,h\,(g,\,L_{(\xi,\delta)}),  \eqno(\z)$$

\noindent where $g=\,  e^{i\,\epsilon ^{\alpha}
P_\alpha}e^{i\,\alpha ^{\mu\nu  }S_{\mu\nu}}
e^{i\,\beta ^{\mu\nu }L_{\mu\nu}}$ and $h:=e^{i\,u ^{\mu\nu }L_{\mu\nu}}$
The nonlinear connection (2.10) adopts now the form

$$\Gamma :=\,L_{(\xi,\delta)}^{-1}\left(d\,+\Omega\,\right)
L_{(\xi,\delta)} =-i\,\vartheta
^\alpha P_\alpha -i\,\Gamma _\alpha {}^\beta \left( S^\alpha {}_\beta
+L^\alpha {}_\beta\right)\,,\eqno(\z)$$

\noindent where $\Omega$ is a one-form affine algebra-valued
with transformation given by (2.9).
\bigskip
\noindent The most outstanding features of this approach are:
\bigskip
i) The connection coefficients can readily be geometrically
identified. The object $\Gamma_\alpha^\beta$ is the geometrical connection
while $\vartheta^\alpha$ represent the tetrads. The essential
advantage of using a nonlinear realization is that, due to
expression (2.11), the transformation of the coframe is the
desired one, i.e. it lacks the inhomogeneous part and the adjoint
piece reproduces the transformation of four Lorentz vectors.
Other approaches proposed to explain the identification of the
coframe and the translational connection recquire the
introduction of mathematical objects which are foreign to the
gauge scheme$^{[1],[8]}$.

ii) There is a correspondence between the mathematical objects of a
standard metric-affine gauge theory$^{[1]}$  and those in
the non-linear gauge version. In addition, the metric tensor,
which is usually introduced by hand in the MAGT, becomes an
intrisic object of the non-linear approach.

iii) We have shown that ordinary matter,
transforming as representations of the Lorentz group, naturally
occurs in the non-linear gauge theories. The embedding of these
physical representations in the linear affine ones is given
by (2.5). In this way we can obtain (reducible) representations
of the Lorentz group starting with the well known tensorial
representations of the affine one and applying the non-linear
procedure. A problem arises when we want to understand how is a
particular spinorial representation of the Lorentz group
imbedded in the linear affine representations. In particular we
will deal with a Dirac spinor. It is natural to expect a
double representation of the Lorentz group to be contained in a
double-valued representation of the affine group. At this
purpose the next logical step is to study the double-valued
representations of the affine group.

\bigskip
\bigskip
\bigskip

\sectio{\bf{Physical spinor representations of $\overline{SL}
$(4,R) }}\bigskip

The existence of spinorial representations for certain linear
groups, such as GL(n, R) or SL(n, R), is well
known by mathematicians. For instance in 1947, Bargmann$^{[9]}$ studied the
double valued representations of SL(2, R). However, for a long time it
was believed by the relativists that there are no spinor fields
transforming linearly
under GL(n, R) or SL (n, R). This wrong belief relied on the
assumption that those groups posess no universal coverings.
Nevertheless, the
existence  of such coverings has been proved by
Ne'eman$^{[10]}$. GL(n, R) and
SL(n, R) have double valued representations, though they come as
infinite matrices.

The systematics of the multipilcity-free unitary irreducible
representations of SL(4,R) were studied by Ne'eman and
{\~{S}}ija{\~{c}}ki$^{[11],[12]}$. They applied their results
 to the description of the phenomenology of hadron
excitations$^{[14]}$. The reader interested in the details of
the derivations reviewed in this section is referred to [1], [11],
[12], [13] and [14].

\bigskip

The affine group GA(4,R) is the semidirect product of the group
of four-dimensional translations and the linear group GL(4,R).
The latter, i.e. the group of invertible $4\times4$ matrices,
contains SL(4,R) as a subgroup , which contains all the
unimodular $4\times4$ matrices , and is a semisimple, non-compact,
non-simply-connected group. Its maximal compact subgroup is
SO(4,R). The feature of one being the maximal compact subgroup
is mantained by their respective covering groups,
 $\overline{SO}(4,R)$ and $\overline{SL}(4,R)$. The former is isomorphic to
SU(2)$\times$SU(2) and it will come apparent the
relevance of this fact when finding the physical spinor representations of
$\overline{SL}(4,\,R)$. As a first question we can ask how is the
Lorentz algebra contained in $\overline{SL}(4,R)$. The traditional
answer is to identify the pseudo-antisymmetric generators of
$\overline{SL}(4,R)$ with the Lorentz ones. Nevertheless, this
approach reveals to be not fully satisfactory.
\bigskip
We now summarize the process of obtention, followed by Ne'eman and
{\~{S}}ija{\~{c}}ki, of all the multipilcity-free
unirreps of $\overline{SL}(4,\,R)$ (unitary, irreducible
representations which, when restricted to the action of
$\overline{SO}(4,R)$,  contain the representation of SU(2)$\times$SU(2)
$\simeq$ $\overline{SO}(4,R)$ labelled by $(j_1,j_2)$ not more than
once). This program is strongly based on the work of
Harish-Chandra$^{[15]}$, and is performed in three steps:

\noindent i) The action of the noncompact generators over the
Hilbert spaces related to the maximal compact subgroup and its
quotient groups is firstly determined.

\noindent ii) Next, the sublattices of the maximal subgroup labels
which are closed under the action of the noncompact generators
are obtained.

\noindent iii) Finally, the constraints on the representation
labels are determined by
imposing a condition of hermiticity on the generators.

\noindent As a result of this procedure, four
series of unirreps are obtained. These unirreps are parametrized
by a complex number $e=\,e_1+ie_2$ and are defined over Hilbert
spaces with values over the quotient groups $K^\prime$ of
SU(2)$\times$SU(2). The scalar product is defined by a kernel
$\kappa$, and the space representation L of $\overline{SL}(4,R)$ is
represented by the smallest values  (min $ j_1$,min $ j_2$) found
between the SU(2)$\times$SU(2) "points" of the sublattice L. The
rest of the points in the sublattice are determined through the
application of the noncompact generators. Therefore, the unirrep
Hilbert spaces are denoted by H($K^\prime ,\kappa,\,L)$.

Besides the trivial representation, the four series of
multiplicity-free unirreps of $\overline{SL}(4,R)$ are

\noindent ---{\it {Principal series}}:  $D^{pr}\,(0,0;e_2)$ and
$D^{pr}\,(1,0;e_2)$,
 $e_1=0, e_2\in R$, defined in the Hilbert spaces
$H(K^{\prime}_1,\kappa,\,L)$ with
$K^{\prime}_1=\,(SU(2)/U(1))\times(SU(2)/U(1)) $,
$\kappa(j_1,j_2)$=1 for all $j_1$ and $j_2$. The irreducible
lattices are respectively L(0,0) and L(1,0).

\noindent ---{\it {Supplementary series}}:  $D^{supp}\,(0,0;e_1)$,
 $0\,< \vert e_1 \vert <\, 1, e_2=0$, defined in the Hilbert spaces
$H(K^{\prime}_1,\kappa,L)$ with
$K^{\prime}_1=\,(SU(2)/U(1))\times(SU(2)/U(1)) $,
$\kappa(j_1,j_2)$ is not trivial, being L(0,0) the irreducible
lattice.

\noindent ---{\it {Discrete series}}:   $D^{dis}\,(1-e_1,0)$ and
$D^{dis}\,(0,1-e_1)$,
 $e_1-j_o, j_o={1 \over{2}},1,{3 \over{2}}, e_2=0$, They are
defined in the Hilbert spaces
$H(K^{\prime}_1,\kappa,L)$ with
$K^{\prime}_1=\,(SU(2)/U(1))\times(SU(2)/U(1)) $. In this case
$\kappa(j_1,j_2)$ is not trivial and the irreducible
lattices are respectively $L(j_o,0;j_1-j_2\ge j_o)$ and
$L(0,j_o,0;j_2-j_1\ge j_o)$.

\noindent ---{\it {Ladder series}}:   $D^{ladd}\,(0,e_2)$ and $D^{ladd}\,
({1 \over{2}},e_2)$,
 $e_1=0, \,e_2\in R$, defined in the Hilbert spaces
$H(K^{\prime}_2,\kappa,L)$ with
$K^{\prime}_2=\,(SU(2)\times SU(2))/SU(2) $,
$\kappa(j_1,j_2)=\kappa(j,j)$=1 for all j;  the irreducible
lattices are respectively L(0) and L(${1\over {2}}$). Now, due to
the form of $K^\prime_2$, the lattices are one-dimensional.

\noindent Let us now discuss the physical interpretation of the
multipilcity-free unirreps of $\overline{SL}(4,R)$.

Starting from a basis of the $\overline{SL}(4,R)$ algebra given by six
pseudo-antisymmetrical
generators $M_{ab}$ and the nine pseudo-symmetrical ones
$T_{ab}$, a,b=0,1,2,3
(remember that one degree of freedom has been removed through
the unimodularity condition), one can define a new basis with
generators

$$\eqalign{&
J_i:=\,{1\over{2}}\epsilon_{ijk}\,M_{jk}\quad ,\quad N_i:=\,
T_{0i}\quad ,\quad K_i:=\,M_{0i}\cr
& T_{ij}\quad ,\quad T_{00}\qquad \,\,\,\,i,j,k=\,1,2,3.\cr}\eqno(\z)$$

\noindent Only $J_i$  and $N_i$ are compact. Among
the most significative subgroups we can identify $\overline
{SO}(4,R)$, generated by $J_i$ and $N_i$, and
$\overline{SO}(3,1)\simeq SL(2,C)$, by $J_i$ and $K_i$.

Using now the  unirrep series of $\overline{
SL}(4,R)$ given above, we observe, that the Lorentz group is
realized by the infinite-dimensional representations of Gel'fand and
Naimark$^{[16]}$, due to the fact that the boost
generators $K_i$ are noncompact. These representations are characterized by the
undesired feature of exciting a spinorial degree of freedom,
i.e. if
we act with $K_i$ over a state of spin j we obtain new states
with spin labels j-1 or j+1. However we know that the actual effect of
applying a boost to a particle is just modifying its momentum.
If the Gel'fand-Naimark representations were really to occur in
Nature, strange situations should be observable:
for instance two inertial observers with relative velocity different
from zero would predict different values for the spin of the
proton. Such unphysical consequences  can be circumvected with the help of
the "deunitarizing automorphism". This automorphism of the
$\overline{SL} (4,\,R)$ algebra, let us call it $A$, acts as

$$\eqalign{{\it{A}}:\,J_i\rightarrow
&\overline{J}_i=\,J_i\quad\,,\quad\,J_i\rightarrow \overline{K}_i=\,i\,N_i
\quad \,,\quad \,T_{00}\rightarrow
\overline{T}_{00}=\,T_{00}\cr
\,N_i\rightarrow &\overline{N}_i=\,i\,K_i\quad\,,\quad\,T_{ij}\rightarrow
\overline{T}_{ij}=\,T_{ij}\quad\,\quad\,i,j=1,2,3.\cr}\eqno(\z)$$

\noindent The same commutation relations hold for both the barred
and the unbarred generators, so that the vector spaces over
which the unirreps of $\overline{SL}(4,\,R)$ are defined are also
available for the realizations of $\overline{SL}(4,\,R)_{{A}}$.
The main difference is that, for the last group, some of the
matrices of the representation are not unitary, due to the $i$
factor in the new generators. As it has been shown by
{\~{S}}ija{\~{c}}ki and Ne`eman$^{[11]}$ this fact does not alter the
hermiticity of the Lagrangian constructed with the new fields,
and we can consider $\overline{J}_i$
and $ \overline{K}_i$ as the physical angular momentum and boost
generators respectively. Under this identification the (finite
dimensional ) representations of the Lorentz group are labelled
by the SU(2)$\times$SU(2) Cassimir labels $(j_1,j_2)$.
\bigskip
Once stated the multipilcity-free unirreps of $\overline
{SL}(4,R)$ and the physical interpretation of its subgroups,
we can consider the question of which of them account for
the fermionic content. With the
identification given above of $\overline
{SO}(4,R)_{{A}}$  as the physical Lorentz subgroup generated by
$\overline{J}_i$
and $ \overline{K}_i$, we impose the
following conditions:

\noindent i) The representation is spinorial, that is,
single-valued for $\overline{SL}(4,\,R)_A$ but double-valued for
${{SL(4,R)}}_A$.

\noindent ii) Certain  "reality conditions" should hold. The states
of the representation must be solutions of an equation
compatible with the equivalence principle$^{[13]}$.
\bigskip
\noindent It is found$^{[13]}$ that the only possible solution
compatible with these constraints leads to a  physical field
belonging to the space of the (reducible)
representation

$$\pi=\,D^{disc}\,( {1\over{2}},0 )_A \oplus D^{disc}\,(
0,{1\over{2}} )_A \eqno(\z)$$

\noindent which is a kind of {\it{extended}} Dirac field.
\bigskip
\bigskip
\sectio{\bf{The non-linear Dirac-affine representation }}\bigskip

Starting from the above given results, we come back to the
problem of how to establish the correspondence between the
non-linear fermionic fields and the linear representation to
which it is related, i.e. we want to apply the results obtained
in the last section to the NLAGT case. To this purpose we must
know the explicit
form of all the objects involved in (2.5).\bigskip
 If we consider the affine non-linear representation $\pi$ and we
want to study its physical content, we need to know how does it
break under the physical Lorentz subgroup $\overline{
SO}(4,R)_{{A}}$. Taking into account the features of the
discrete series given in the last section, it is easily
found that the "Lorentz content" of $\pi$ is given by

$$\{(j_1,j_2)\}=\,\{({1 \over{2}},0),({3 \over{2}},1),({5 \over{2}},2),...\}
\oplus \{(0,{1 \over{2}}),(1,{3 \over{2}}),(2,{5 \over{2}}),...\}.
\eqno(\z)$$

\noindent The effect of the non-compact generators is to mix the different
spin subspaces. For instance, if we start with a pure state
which only projects over the $({1 \over{2}},0)\oplus (0,{1
\over{2}})$ subspace, once we act on it with the non-compact
generators ( which, as we
shall see, are related to the symmetric
ones of the affine group), becomes a superposition of new spin
states. So, we obtain that the action of SL(4,R) over its
representation states reproduces a particle creation-like situation.
This phenomenon has been studied$^{[14]}$ in the QCD context,
trying to explain the spectrum of the hadronic excitations.
Nevertheless, in a gauge theory of gravitation, the particle
creation might be considered as an undesired property. We
will see how the nonlinear approach removes it.\bigskip

We now apply the non-linear procedure described in section 2 to
the ${{SL(4,R)}}_{{A}}$ unirreps of section 3. At first sight,
there is an incompability between the non-linear affine
realizations studied in [2] and [3] and the approach of section
3, where only the SL(4, R) unirreps were found. A first solution
is obtained when we consider the non-linear realization of SL(4,
R) with the Lorentz subgroup as the classification one. This solution is not
entirely satisfactory, since SL(4, R) lacks the translational part
which, as shown in [3], is required in the context of a gauge
theory of gravitation in order to identify the coframe with the
translational connection. We now see how the SL(4, R)
representations can be extended to the whole affine group. To
this purpose we define the group $F_4$ as the semidirect product of
the four dimensional translations and the uniparametric group of
the four dimensional matrices proportional to the unity. The two
following properties can be easily proved \bigskip

\noindent $F1$)  $F_4$ is a normal subgroup of the affine group.

\noindent $F2$) Every element $g$ in the affine group can be uniquely written
as $g=fs$ where $f\,\in F_4$ and $s\in\,$SL(4,R).

Starting from a representation $\sigma(s),\,s\in\,$SL(4,R), we can
define the application $\Sigma$ defined over the affine group through

$$\Sigma(g):=\,\sigma(s)\quad\,;\qquad g=fs\quad,\quad\,f\in F_4
\quad,\quad\,s \in SL(4,R).
\eqno(\z)$$

\noindent $\Sigma (g)$ is well defined over the affine group
because of $F2$. Let us now check, using $F1$ and the homomorphism
property of $\sigma$, that $\Sigma$ is by itself a representation:

$$\Sigma(g_1\,g_2)=\,\Sigma(f_1s_1\,f_2s_2){\buildrel{F_1}\over
{=}}\, \Sigma(f_1f_3s_1s_2)=
 \Sigma(f_4s_1s_2)=\,\sigma(s_1)\,\sigma(s_2)=\,\Sigma(g_1)\,\Sigma(g_2)$$

\noindent Hence, starting from (4.2), we can obtain a representation of
the affine group once a  representation of SL(4,R) is
known. Using (3.3) we now define

$$\Delta(g):=\,\pi(s)\quad ;\quad g=fs\quad ,\quad\,f\in F_4
\quad,\quad \,s \in SL(4,R),
\eqno(\z)$$

\noindent which is a representation of the affine group and will
play the role of $\varphi$ in (2.1) and (2.5) when
applying the non-linear procedure. As a first step we work out
its breaking under the Lorentz group given by (4.1).
\noindent We wish to label each one of the (infinitely many) spaces appearing
in (4.1) by an index i=1,2,3...
. To do so we write an infinite matrix where the first
row corresponds to the $({1\over{2}},0)$ subspace, the second to
to the $({3\over{2}},1)$ and so on. Correspondingly, the first
column is related to the $(0,{1\over{2}})$ subspace, the second
to the $(1,{3\over{2}})$, etc. The direct sum of the
infinite elements of the matrix yields the space of
representation (4.1) of $\Delta$. Starting from the
first element in the first row, we get through all the matrix
elements by following the alignements of slope +1 from lower
left to upper right. All the subspaces occurring in (4.1) are
thus pinpointed. In this way
we obtain

$$\eqalign{\{(j_1,j_2)\oplus (j^\prime_1,j^\prime_2)\}=&\{
({1\over{2}},0)\oplus(0,{1\over{2}}),({1\over{2}},0)\oplus(1,
{3\over{2}}),({3\over{2}},1)\oplus(0,{1\over{2}}),\cr
({1\over{2}},0)&\oplus(2,{5\over{2}}),({3\over{2}},1)\oplus(1,
{3\over{2}}),({5\over{2}},2)\oplus(0,{1\over{2}}), ...
\}.\cr}\eqno(\z)$$

\noindent The number of the places occupied in (4.4) labels
the number of the subspace $V_i$, i=1,2,3,... , defined as the
subspaces into which the representation $\pi$ (and therefore
$\Delta$, because of (4.3)) breaks under the action of the
physical Lorentz group. The basis of the
representation V of $\Delta$ can be chosen in such a way that

$$V=\,\oplus\sum_{i=1}^\infty \,V_i\,\quad,\eqno(\z) $$

\noindent and $\Delta$ becomes

$$\Delta=\left(\matrix{\Delta_1\cr
                       \Delta_2\cr
                       \Delta_3\cr
                        ...\cr}\right).\eqno(\z)$$

\noindent Its transformation under the
action of an element g of the affine group, induced by the variation of $\pi$,
see (4.3), is of the following form

$$\Delta\rightarrow\Delta^\prime=\left(\matrix{\Delta_1^\prime\cr
                       \Delta_2^\prime\cr
                        ...\cr}\right)=
                \left(\matrix{D_{11}(g)&D_{12}(g)&...\cr
                              D_{21}(g)&D_{22}(g)&...\cr
                              ...&       ...     &...\cr}\right)
                     \left(\matrix{\Delta_1\cr
                       \Delta_2\cr
                       ...\cr}\right),\eqno(\z)$$

\noindent or briefly

$$\Delta_i^\prime=\sum_{j=1}^\infty
D_{ij}(g)\,\Delta_j\,. \eqno(\z)$$

\noindent Now, following eq. (2.5), we find
the non-linear partner of $\Delta$. The procedure was already
pointed out by us choosing the coset structure
(2.14). One point must be first clarified. In section 3 we saw
that the Lorentz algebra was implemented by the so(4,R)$_A$
subalgebra of sl(4,R)$_A$. So, we have to identify the
generators used in [2] and [3] with those of sl(4,R) through

$$\eqalign{\overline{J}_i=&\,{1\over{2}}\epsilon_{ijk}L_{jk}\cr
           \overline{K}_i=&\,iS_{0i}\cr
\overline{N}_i=&\,iL_{oi}\,\,\,\,\,\,i,j,k=1,2,3\cr}\eqno(\z)$$

\noindent plus the remaining pseudosymmetrical ones. The
remaining generators needed to obtain the affine
algebra, departing from sl(4,R)$_A$, are taken from $F_4$.
Following the previous identification we define

$$\phi:=\,D({L_{(\xi,\delta)}}^{-1})\,\Delta \eqno(\z)$$

\noindent where D is given by (4.7) and ${L_{(\xi,\delta)}} $
by (2.14). Making use of (2.6) and knowing the Lorentz
decomposition of $\Delta$ (4.5), the
transformation of
$\phi$ under the action of an element g of the affine group is
easily derived, namely

$$\phi\rightarrow\phi^\prime=\left(\matrix{\phi_1^\prime\cr
                       \phi_2^\prime\cr
                        ...\cr}\right)=
                \left(\matrix{D_{11}(h)&0&...\cr
                              0&D_{22}(h)&...\cr
                              ...&       ...     &...\cr}\right)
                     \left(\matrix{\phi_1\cr
                       \phi_2\cr
                       ...\cr}\right),\eqno(\z)$$

\noindent with $h=\,h(g,\,{L_{(\xi,\delta)}}$ ) given by eq. (2.15).

\noindent Eq.(4.9) can be rewritten in shorthand as

$$\phi_i^\prime=\,
D_{ii}(h(g,\,{L_{(\xi,\delta)}}))\,\phi_i,\eqno(\z)$$

\noindent and shows that the non-linear action of the affine
group does not mix the different spin subspaces. This is
easy to understand since, through the non-linear realization,
all the linear generators of the affine group are non-linearly projected over
the generators of the Lorentz group.\bigskip

If we focus our attention on the $\phi_1$ component of $\phi$,
we realize that it transforms according to the $({1\over{2}},0) \oplus
(0,{1\over{2}})$ Lorentz representation with non-linear coefficients given by
(2.15). It does not get mixed with other spin states under the
action of the affine group, and behaves as usual under the
Lorentz group (see (2.3)). It then becomes the non-linear
affine version of the Dirac field
in the context of the NLAGT.

\bigskip
\bigskip
\sectio{\bf{Conclusions}}\bigskip
 Following the lines set in [2] and [3], we have obtained
an infinite-dimensional non-linear realization of the affine
group which reproduces a fermionic spectrum. Particle
creation-like effect is absent, due to the special properties of
the non-linear realizations. The $\phi_1$ component can be
identified with the Dirac field transforming as a $({1\over{2}},0) \oplus
(0,{1\over{2}})$ Lorentz representation with non-linear
coefficients. This enlightens the inclusion of ordinary fermionic
matter in the affine gauge theories of gravitation and can be
considered as a strong argument in favour of the non-linear
approaches.  \bigskip\bigskip
\noindent{\bf Acknoledgement}\vskip1.0cm
\noindent We are very grateful to Dr Jaime Julve for helpful
discussions. \bigskip\bigskip

\centerline{REFERENCES}\vskip1.0cm

\noindent\item{[1]} F. W. Hehl, J. D. McCrea, E. W. Mielke and
Y. Ne'eman, {\it Phys. Rep.} {\bf 258} (1995) 1

\noindent\item{[2]}   A. L\'opez-Pinto, A. Tiemblo and R. Tresguerres,
{\it Class. Quantum Grav.} {\bf  12} (1995) 1503

\noindent\item{[3]}   J. Julve, A. L\'opez-Pinto, A. Tiemblo and R.
Tresguerres,
to appear in {\it New Frontiers in Gravitation,} G. Sardanashvily
and R. Santilli eds, Hadronic Press, Inc., Palm Harbor (1995)

\noindent\item{[4]} L.N. Chang and F. Mansouri, {\it Phys. Rev.}
{\bf D 17} (1978) 3168

\noindent\item{[5]} F. Grsey and L. Marchildon, {\it Phys. Rev.}
{\bf D 17} (1978) 2038

\noindent\item{[6]} S. Coleman, J. Wess and B. Zumino, {\it Phys. Rev.}
{\bf 177} (1969) 2239

\noindent\item{[7]} A. Salam and J. Strathdee, {\it Phys. Rev.}
{\bf 184} (1969) 1750

\noindent\item{[8]} E. W.  Mielke, J. D. McRea, Y. Ne'eman and F.
W. Hehl, {\it Phys. Rev.} {\bf D 48} (1993) 673 and references therein

\noindent\item{[9]} V. Bargmann, {\it Ann. of Math.}
{\bf 48} (1947) 568

\noindent\item{[10]} Y. Ne'eman, {\it Ann. Inst. Henri Poincar'}
{\bf A 28} (1978) 369

\noindent\item{[11]}   Dj. {\~{S}}ija{\~{c}}ki and Y. Ne'eman,
{\it J.Math.Phys.} {\bf  26} (1985) 2457

\noindent\item{[12]} Y. Ne'eman and Dj. {\~{S}}ija{\~{c}}ki, {\it
Ann. of Phys.} {\bf 120} (1979) 292

\noindent\item{[13]}   A. Cant and Y. Ne'eman,
{\it J.Math.Phys.} {\bf  26} (1985) 3180

\noindent\item{[14]} Y. Ne'eman and Dj. {\~{S}}ija{\~{c}}ki, {\it Phys. Rev.}
{\bf D 37} (1988) 3267

\noindent\item{[15]} Harish-Chandra, {\it Proc. Nat. Acad. Sci. U.S.A.}
{\bf 37} (1951) 170, 362, 366, 691

\noindent\item{[16]} I.M. Gel'fand and M.A. Naimark, {\it Izv.
Akad. Nauk. SSSR, Ser. Mat.}
{\bf 11} (1947) 411

\bye